\documentclass[runningheads]{llncs}
\usepackage[T1]{fontenc}
\usepackage{graphicx}
\usepackage{amsmath}
\usepackage{amsfonts}
\usepackage{subcaption}
\usepackage{booktabs}
\usepackage{multirow}

\begin{document}
\title{Five Pitfalls When Assessing Synthetic Medical Images with Reference Metrics}
\titlerunning{Five Pitfalls of Reference Metrics in Medical Imaging}
%
\author{Melanie Dohmen, Tuan Truong, Ivo M. Baltruschat, Matthias Lenga}
%
\authorrunning{Dohmen et al.}
%
\institute{Bayer AG, Berlin, Germany\\
\email{firstname.lastname@bayer.com}\\}
\maketitle              
\begin{abstract}
Reference metrics have been developed to objectively and quantitatively compare two images. Especially for evaluating the quality of reconstructed or compressed images, these metrics have shown very useful. Extensive tests of such metrics on benchmarks of artificially distorted natural images have revealed which metric best correlate with human perception of quality. Direct transfer of these metrics to the evaluation of generative models in medical imaging, however, can easily lead to pitfalls, because assumptions about image content, image data format and image interpretation are often very different. Also, the correlation of reference metrics and human perception of quality can vary strongly for different kinds of distortions and commonly used metrics, such as SSIM, PSNR and MAE are not the best choice for all situations. We selected five pitfalls that showcase unexpected and probably undesired reference metric scores and discuss strategies to avoid them.

\keywords{reference metrics, image synthesis, similarity, normalization}
\end{abstract}
\section{Introduction}

A large set of image reference metrics has been developed for the assessment of image compression and image reconstruction algorithms. The Tampere Image Database \cite{tid_dataset} and the LIVE Image Quality Assessment Database \cite{full_reference_metrics,LIVE} are frequently used benchmark datasets including human quality assessments of artificially distorted images to identify those reference metrics, that best correlate with human perception across all distortion types. Predominantly, noise (additive, impulse, block-wise etc.) and JPEG compression artifacts are included. Also, a similar study with magnetic resonance (MR) images \cite{SubjectiveAssessmentMR} evaluated reference metrics regarding their sensitivity regarding mainly noise and compression artifacts. Although these studies aim for finding a single metric equally sensitive to all distortions, the results clearly suggest that different metrics perform best with certain distortion types.
According to a review on generative adversarial networks (GANs)\cite{MacNaughton2023} for image-to-image translation in medical images, the most frequently used metrics are mean absolute error (MAE), the structural similarity index measure (SSIM)\cite{ssim}, and the peak signal-to-noise ratio (PSNR)\cite{psnr}, even though PSNR and MAE were shown to badly correlate with human perception and SSIM was found to perform especially well in the group of JPEG compression artifacts \cite{tid_dataset}. If these metrics are really appropriate for assessing the quality of synthetic medical images is at least questionable. Finding metrics for evaluation goes inline with finding loss functions for  model training. In this context, learned metrics have shown more suitable \cite{ding2020optim}.

In this paper, we want to showcase and explain five pitfalls, that we have observed when evaluating synthetic medical images with different kinds of reference metrics. Some of them relate to specific distortions that are not commonly tested with reference metrics. Other potential pitfalls arise from different data formats in medical images compared to natural images and that image content and interpretation are more important in the medical domain.
In comparison to previous work \cite{metric_pitfalls} on metric related pitfalls, which includes primarily segmentation metrics, this study focuses on reference metrics, that measure the similarity directly between two images.

\section{Data and Methods}
\subsection{Normalization and Binning}
Normalization aims to shift and rescale the intensity range of an image $I$ to make it better comparable to the intensity range of another image.
\begin{equation}
I^{\prime} = (I - a ) / b
\label{eq:norm}
\end{equation}
Using the minimum intensity $a=I_{\mathrm{min}}$ and the difference of maximum and minimum intensity $b=I_{\mathrm{max}}-I_{\mathrm{min}}$, normalization is often referred to as Minmax normalization. When normalizing with the mean $a=\mu$ and the standard deviation $b=\sigma$, normalization is often referred to as Zscore normalization.
To transform images with higher intensity ranges into 8-bit integer format, commonly binning with $b=256$ bins is used:
\begin{equation}
I^{\prime}
= \min(b-1,
\lfloor(I - I_{\mathrm{min}}) / b \cdot(I_{\mathrm{max}}-I_{\mathrm{min}}) \rfloor )
\label{eq:binning}
\end{equation}
\subsection{Reference Metrics}
\label{sec:metrics}
Given a reference image $R$ and a test image $I$ of the same width $w$, height $h$ and if applicable depth $d$, a reference metric $m(R,I): \rightarrow \mathbb{R}$ assigns a real-valued score.
Among the most popular reference metrics is SSIM\cite{ssim}, which compares contrast, mean intensity and structure in a local sliding window between the test image and the reference image. Its multi-scale variant MS-SSIM\cite{msssim} calculates and combines multiple scores for different downscaled versions of the images. SSIM is parametrized with a data range parameter $L$, which depends on the intensity value range of the images. The default is 255 for 8-bit images. For any other data format, we propose the joint range of both images, but $L$ must be chosen with care (see Sec.\,\ref{sec:pitfall1}).
Another variant of SSIM, calculated on complex-wavelet transformed images called CW-SSIM, does not include a parameter $L$ and was proposed to be less sensitive to small rotations, scale or translation \cite{cw-ssim}.
A group of error metrics including mean absolute error (MAE) and mean squared error (MSE) directly depend on the differences of intensity values at all 
 $N = w \cdot h (\cdot d)$ corresponding pixel locations $\mathbf{x}$ in $I$ and $R$.
The peak-signal-to-noise-ratio (PSNR)\cite{psnr} is defined via the MSE and also parametrized by a data range parameter $L$. As for SSIM, $L=255$ has been proposed for 8-bit integer data, while we use $L = \mathrm{max}(I_{\mathrm{max}}, R_{\mathrm{max}}) - \mathrm{min}(I_{\mathrm{min}}, R_{\mathrm{min}})$ for all other intensity ranges as default.
Another group of learned metrics is based on features extracted by pre-trained classification networks. The Learned Perceptual Image Patch Similarity (LPIPS) additionally weights these features for optimal similarity judgement. Deep Image Structure and Texture Similarity (DISTS) adapts the LPIPS metrics by varying network elements, weighting factors and feature comparison to be more sensitive to texture similarities. Learned metrics depend on the trained network in terms of training data and architecture. For LPIPS as a forward metric, the AlexNet backbone is recommended\cite{lpipspypi}.
A further group of metrics quantifies the degree of statistical dependency of images $I$ and $R$. The Pearson Correlation Coefficient \cite{PET2CT} measures the degree of linear dependency between the pixel intensities $I(\mathbf{x})$ and $R(\mathbf{x})$ for all pixel locations $\mathbf{x}$. Mutual information (MI) \cite{mutual_info} sums the entropies of $I$ and $R$ and subtracts the joint entropy. Normalized mutual information (NMI)\cite{skimage} divides by the joint entropy instead of subtracting it.

The task-specific similarity of $I$ and $R$ can also be compared after performing a downstream task with $I$ and $R$ and assessing the similarity of the results. A common downstream task is segmentation, and then the segmentation results of $I$ and $R$ are evaluated by a segmentation metric. A very popular segmentation metric is the DICE \cite{dice_original} score.
The evaluation of a metric can be restricted to certain pixel locations $\mathbf{x}$. For example, a mask could indicate background and only include pixels in the foreground for the calculation of the metric score. 
When error metrics are calculated from pixel intensities at single locations, such as all error metrics and statistical dependency metrics, masking is easily applicable. However, if metrics are not calculated from each pixel location separately, and combine information from neighboring pixel locations, such as SSIM, MS-SSIM, CW-SSIM, LPIPS or DISTS, masking with non-rectangular masks is not directly applicable. Masking with rectangular masks is basically equivalent to evaluating the mask on cropped images.
\section{Experiments and Results}
All experiments were performed with T1-weighted contrast-enhanced MR images from the first 100 cases of the BraSyn Dataset \cite{BraSyn}. We apply certain normalizations (default: no normalization) or distortions and evaluate all introduced metrics (SSIM, MS-SSIM, CW-SSIM, PSNR, MAE, MSE, LPIPS, DISTS, NMI and PCC) in order to uncover important differences.
\begin{figure}[htb]
\includegraphics[width=\textwidth]{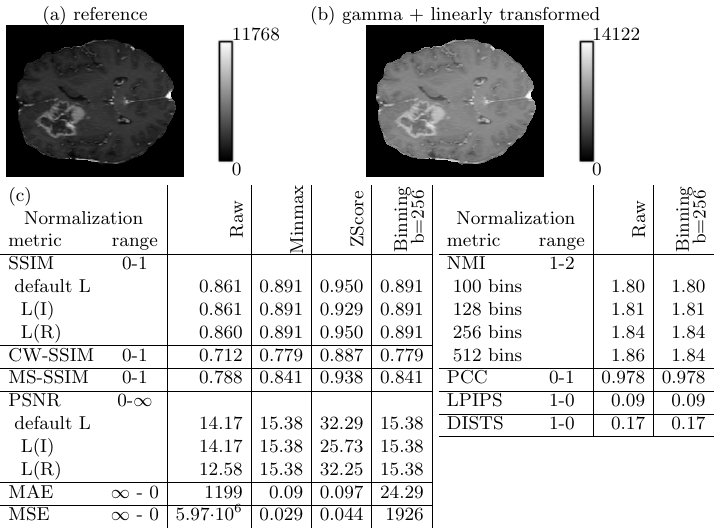}
\caption{An example reference image (a) and its gamma and linearly transformed version (b) are shown. Mean similarity scores over 100 images are listed in (c). The results reveal strong influence of normalization parameters and methods.}
\label{fig:pitfall1}
\end{figure}
\subsection{Pitfall 1: Inappropriate Normalization}\label{sec:pitfall1}
Challenges with normalization arise, when the intensity value ranges of two images $I$ and $R$ are not equal. When shape of the histograms of $I$ and $R$ are not alike, metric evaluation after Zscore normalization may deviate from evaluation after Minmax normalization. Also, small deviations of the data range parameter $L$ for SSIM and PSNR may have a noticeable effect on the metric scores.

We distorted the images by applying a gamma transform with $\gamma=0.4$ and subsequent linear scaling with $f=1.2$ and calculate the metrics in comparison to the undistorted original images.
We evaluated SSIM and PSNR with the default data range $L = \mathrm{max}(I_{\mathrm{max}}, R_{\mathrm{max}}) - \mathrm{min}(I_{\mathrm{min}}, R_{\mathrm{min}})$ (see Sec.\,\ref{sec:metrics}), but also with $L(I) = I_{\mathrm{max}}-I_{\mathrm{min}}$ and $L(R) = R_{\mathrm{max}}-R_{\mathrm{min}}$. We evaluated all metrics with Minmax, Zscore and without normalization as well as with binning to 256 bins (see Eq.\,\ref{eq:binning}). As LPIPS and DISTS require images with an intensity range fixed to $[-1, 1 ]$ and $[0,1]$ respectively, default application includes a normalization according to Eq.\,\ref{eq:norm} with $a=(I_{\mathrm{min}}+I_{\mathrm{max}})/2$ and $b=(I_{\mathrm{max}} -I_{\mathrm{min}})/2$ and Minmax normalization respectively. Therefore, the learned metrics are not additionally evaluated with Minmax and Zscore normalization. PCC and NMI are, by definition (see Sec.\,\ref{sec:metrics}), not sensitive to normalization as defined in Eq.\,(\ref{eq:norm}) and are also not evaluated with Minmax and Zscore normalization. However, we evaluate NMI for internal binning with 128, 256 and 512 bins.

The results in Fig.\,\ref{fig:pitfall1} show that SSIM and PSNR increase for higher data ranges and decrease for binned data. Zscore and Minmax normalization result in noticeably different metric scores, because reference and transformed images have different intensity ranges and different means. NMI almost ignores the gamma and linear transform, when internally a high bin number is used and binning was not performed as pre-normalization. Smaller bin numbers and non-matching internal and pre-binning bin numbers may further reduce similarity artificially.
\subsection{Pitfall 2: Similarity of Misaligned Images}
\begin{figure}[htb]
\includegraphics[width=\textwidth]{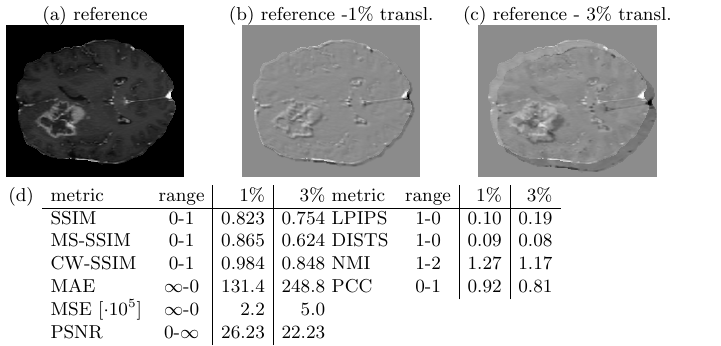}
\caption{Small misalignments have strong influence to all reference metrics. Only DISTS and CW-SSIM are less sensitive to small geometric transformations.}
\label{fig:pitfall2}
\end{figure}

In image-to-image translation, the source domain input image and the target domain image are often misaligned, because both images were acquired at different timepoints or even with different devices. Therefore, an image synthesized from a misaligned input image is also often misaligned. However, in most cases, medical images are perceived as similar and interpreted in the same way, regardless of small spatial misalignments. Fig.\,\ref{fig:pitfall2} shows that small translations, that are hardly visible, significantly affect most metric scores. Only CW-SSIM and DISTS do not show large changes as they were designed and reported to be less sensitive to misaligments. 
\begin{figure}[htb]
\includegraphics[width=\textwidth]{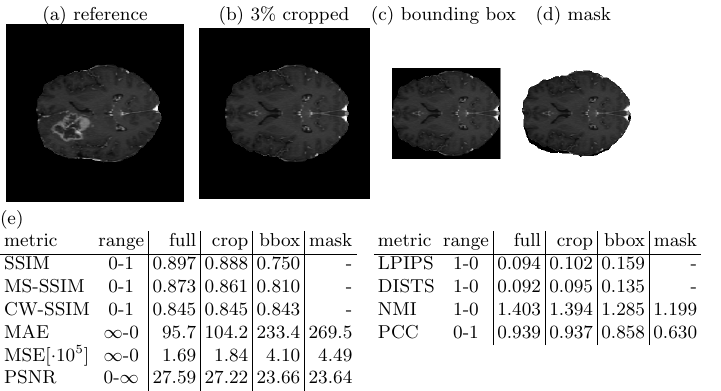}
\caption{An example reference image (a), its by 3\% cropped version (b), its by a bounded box cropped version (c) and an exactly foreground masking version (d) are shown. Mean similarity scores over 100 images are listed in (e). With less identical background included in the calculation, the assessed similarity strongly decreases.}
\label{fig:pitfall3}
\end{figure}
\subsection{Pitfall 3: Background, Foreground and Region of Interest Similarity}
Medical images are often acquired to detect a pathological condition in a very specific location in the human body. Even though the field of view can be narrowed, medical images often picture neighboring structures and a large fraction of background, that are not of interest for diagnosis. Similarity of medical images is especially relevant for a limited region of interest, i.e. a possible lesion or tumor, a specific organ, bone, muscle or tendon. Pictures of brain tumors are perceived more similar, if they show the same type of tumor at the same location, rather than the same texture of healthy brain tissue or even the same background intensity. Therefore, it is important to be able to mask out rather irrelevant parts of an image and to evaluate specified regions of interest separately. Fig.\,\ref{fig:pitfall3} shows similarity metric scores for increasingly cropped brain images, where the test image consists of the upper hemisphere of the brain in the reference image and the lower half of the image is replaced by a mirror of the upper hemisphere. In most cases of the BraSyn Dataset this leads to either two or no tumors in the test image opposed to exactly one tumor in the reference image. However, when background composes a large part of the image to be evaluated, similarity metric scores appear very high. With decreasing background the similarity metric scores also noticeably decrease.
\subsection{Pitfall 4: Error Metrics Prefer Blurred Images}
\begin{figure}[htb]
\includegraphics[width=\textwidth]{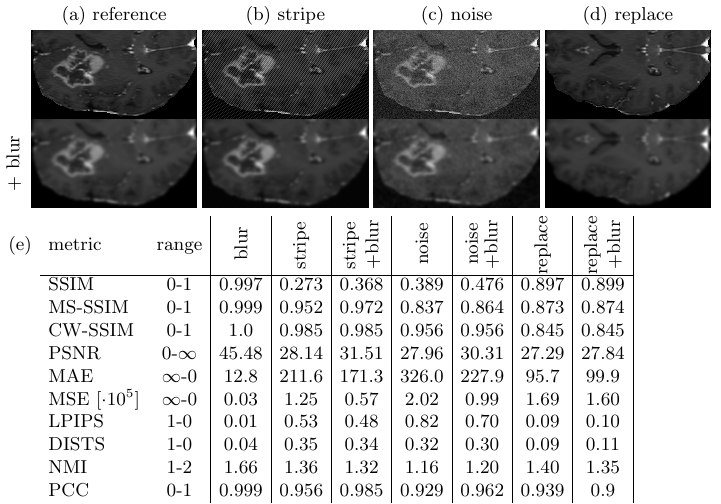}
\caption{An example region of interest with different distortions is shown in the first row: (a) reference, (b) stripes added, (c) Gaussian noise added, (d) lower half of the image replaced by mirror of the upper half. Mean similarity scores were assessed over 100 images (e). Blurring perceptually improves strong distortions and quantitatively improves most similarity scores, especially SSIM. Out of all observed metrics, NMI best detects blurring.}
\label{fig:pitfall4}
\end{figure}

When using loss functions based on error metrics, such as MSE, it has been reported and observed, that optimized models generate blurry images \cite{DyeFreeNet2020}. We assessed metric scores for three kinds of distortions and also for the undistorted images with additional blurring. We observe, that metric scores increase for the additionally blurred versions. The distortions are also perceived as weakened by the blurring. However, the overall quality and degree of blurriness is not satisfactory and we assume further blurring will not arbitrarily improve similarity. The metric score results and example images are shown in Fig.\,\ref{fig:pitfall4}.
\subsection{Pitfall 5: Perceptual and Task-Specific Similarity}
\begin{figure}[htb]
\includegraphics[width=\textwidth]{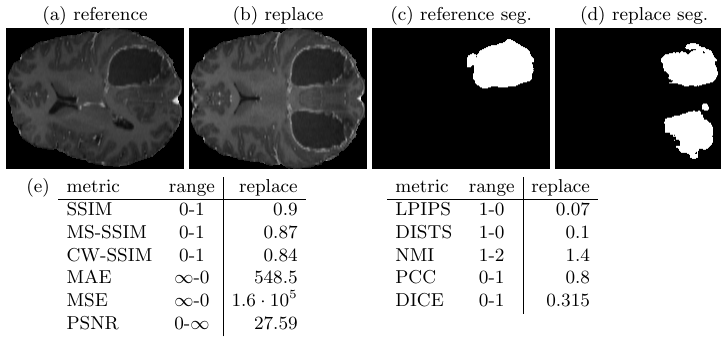}
\caption{An example of a reference image (a) and a version with replacements (b), as well as their respective tumor segmentations (c, d) are shown. Specifically, the lower half of the reference images are replaced by the mirrored upper half.  The mean similarity scores over 100 images are assessed by different metrics (e). While most similarity metrics hardly change with artificial introduction or removal of a tumor, additional or missing tumor segmentations strongly decrease the DICE score.}
\label{fig:pitfall5}
\end{figure}

Similar to the masking, for medical imaging, a possible tumor is probably one of the most important structures to be correctly synthesized in an MR image of the human brain. However, if there is only a mask of the tumor in the reference image, artificially synthesized tumors in healthy tissue regions in the synthetic image are easily overlooked. Tumors may also be very heterogeneous in their texture and local structure, such that similarity metrics restricted to the tumor region are not informative about the similarity of the tumor type. Therefore, it can be useful to define and perform an important downstream task with the synthetic images. Then the similarity of the synthetic images to the reference images can be assessed by comparing the performance of the downstream task results of both image subsets. If both images lead to very similar results, the synthetic and the reference image appear similar regarding the tested task. 
In this case, we trained an automatically configuring U-Net based segmentation network \cite{monai_autoseg,seg_model} on the T1c images of the BraSyn dataset\cite{BraSyn} and the whole tumor annotations.
The architecture of the U-Net included five residual blocks, with downsampling factors 1, 2, 2, 4 and 4, initially 32 features and one output channel activated by a sigmoid function. As a preprocessing step for training and inference, Zscore normalization was applied to the input images. 
In Fig.\,\ref{fig:pitfall5} example segmentations are shown and in addition to the previous metrics, the DICE score was assessed from the segmentation results. Especially compared to SSIM, the extra or missing tumors are clearly detected by the DICE score.
\section{Discussion and Conclusion}
In this study we have shown that many types of reference metrics exists with different sensitivity to normalization, misalignment, blurring and masking. Most of these metrics, including SSIM and PSNR, were first developed and designed for 8-bit integer valued natural images for quantifying image quality after image compression or reconstruction. However, now they are often used for assessing similarity between synthetic medical images and real medical reference images. Although SSIM correlates well with human perception, there are specific distortions such as blurring or replace artifacts, where other metrics, such as LPIPS, NMI or a downstream segmentation metric are more appropriate and should be additionally considered. When working with non 8-bit integer images, especially normalization and binning of float-valued data formats must be performed with care and all parameters must be documented, because slight differences have high impact. Further, we showed that misalignment of multi-modal data, that is often used for image synthesis, may impair evaluation. Better pre-registration or the selection of suitable image metrics such as CW-SSIM or DISTS are possible solutions.
The use of non-reference metrics, which were shown to detect typical distortions of medical images \cite{dohmen2024similaritymetricsmrimagetoimage}, could approach potential issues with unpaired data.
At last, the interpretation of image content plays a central role in the medical domain. Knowledge about regions of interest, downstream detection, segmentation or classification tasks can and should be used to evaluate task-specific similarity.
In summary, we recommend to carefully consider the type of expected distortions in the image domain and to select a suitable set of reference metrics. Proper registration, normalization and masking of regions of interest additionally improve the reliability, when evaluating synthetic medical images.
\begin{credits}
\subsubsection{\discintname}{The authors have no competing interests to declare that are relevant to the content of this article.}
\end{credits}
\bibliographystyle{ieeetr}
\bibliography{Paper-0021}

\end{document}